\documentclass[pre, 
aps, 
amssymb, 
amsmath, 
nofootinbib, 
twocolumn, 
superscriptaddress]{revtex4-1}

\usepackage{graphicx}
\usepackage{color}
\usepackage{verbatim}
\usepackage{subfigure}
\usepackage{amssymb}
\usepackage{mathtools}
\usepackage{romannum}

\newcommand{\pot}{V_{\boldsymbol{\lambda}(t)}}
\newcommand{\Green}{G_{\boldsymbol{\lambda}(t)}}
\newcommand{\FPop}{\hat{\mathcal{L}}_{\boldsymbol{\lambda}(t)}}
\newcommand{\adjointFPop}{\hat{\mathcal{L}}^{\dagger}_{\boldsymbol{\lambda}(t)}}

\newcommand{\rhoeq}{\rho^{eq}_{\boldsymbol{\lambda}(t)}}

\newcommand{\Schr}{\hat{\mathcal{H}}_{\mathcal{S}}}
\newcommand{\Schrpot}{U_{\boldsymbol{\lambda}(t)}}
\newcommand{\Sym}{\hat{\mathcal{H}}}
\newcommand{\metric}{\boldsymbol{\zeta}(\boldsymbol{\lambda})}
\newcommand{\SchrGreen}{G_{\mathcal{S}}}

\newcommand{\FPopshort}{\hat{\mathcal{L}}}
\newcommand{\rhoeqshort}{\rho^{eq}}

\newcommand{\pieq}{\Pi^{eq}_{\boldsymbol{\lambda}(t)}}

\definecolor{purple}{rgb}{0.5,0,0.5}
\definecolor{darkgreen}{rgb}{0,0.5,0}
\definecolor{orange}{rgb}{0.8,0.33,0}

\DeclareMathOperator*{\argmin}{arg\,min}

\begin{document}

\title{Solution to the Fokker-Planck equation for slowly driven Brownian motion: Emergent geometry and a formula for the corresponding thermodynamic metric}

\author{Neha S. Wadia}
\email{neha.wadia@berkeley.edu}
\affiliation{Biophysics Graduate Group, 
University of California at Berkeley, Berkeley, California 94720, USA}
\author{Ryan V. Zarcone}
\affiliation{Biophysics Graduate Group, 
University of California at Berkeley, Berkeley, California 94720, USA}
\author{Michael R. DeWeese}
\affiliation{Biophysics Graduate Group, 
University of California at Berkeley, Berkeley, California 94720, USA}
\affiliation{Department of Physics, 
University of California at Berkeley, Berkeley, California 94720, USA}
\affiliation{Redwood Center for Theoretical Neuroscience 
and Helen Wills Neuroscience Institute, 
University of California at Berkeley, Berkeley, CA 94720, USA}

\begin{abstract}
Considerable progress has recently been made with
geometrical approaches to understanding and controlling small out-of-equilibrium 
systems, but a mathematically 
rigorous foundation for these methods has been lacking. 
Towards this end, we develop a perturbative solution to the Fokker-Planck equation for 
one-dimensional driven Brownian motion in the overdamped limit 
enabled by the spectral properties of the corresponding single-particle
Schr\"{o}dinger operator.
The perturbation theory is in powers of the inverse characteristic timescale
of variation of the fastest varying control parameter, measured in units of the system timescale, 
which is set by the smallest eigenvalue of the corresponding Schr\"{o}dinger operator.
It applies to
any Brownian system for which the Schr\"{o}dinger operator has 
a confining potential.
We use the theory to rigorously derive an exact formula
for a Riemannian ``thermodynamic" metric in the space of control parameters of the system.
We show that up to second-order terms in the perturbation theory, optimal 
dissipation-minimizing driving protocols minimize the length defined by this metric.
We also show that a previously proposed metric 
is calculable from our exact formula with corrections that are exponentially suppressed in a 
characteristic length scale.
We illustrate our formula using the two-dimensional example of a harmonic oscillator with time-dependent 
spring constant in a time-dependent electric field.
Lastly, we demonstrate that the Riemannian geometric structure of the optimal control problem 
is emergent; it derives from the form of the perturbative expansion for the probability density 
and persists to all orders of the expansion.
\end{abstract}

\maketitle

Driven Brownian motion is a paradigmatic model for a certain class of small 
(micrometer sized and smaller) stochastic machines~\cite{seifert-review-2012}.
The hallmark of these systems is that important quantities such as work and 
efficiency fluctuate, and are comparable in scale to thermal fluctuations.
Their study, i.e., stochastic thermodynamics~\cite{sekimoto2010}, 
has seen remarkable recent experimental progress
\cite{
liphardt2002,  % experimental test of jarzynski inequality
trepagnier2004-hatano-sasa-experiment,  % experimental test of hatano-sasa inequality
collin2005-RNA-folding, % test of crooks fluctuation theorem
toyabe2010, % information-to-heat engine
mehl2010, % FDT for driven Brownian particle
gomez-solano2011,
berut2012-experimental-landauer, % verification of landauer principle
berut2013,
quinto-su2014,
koski2014,
roldan2014,
jun2014-high-precision-landauer, % landauer/jarzynski
dieterich2015, % single-molecule measurement of effective temperature
hong2016, % landauer limit
gavrilov2016-PRL, % bit erasure/landauer/distinction between thermodynamic and logical reversibility
frey2015-DNA-stretching, % free energy landscape of some DNA molecule
martinez2015-trapped-brownian-particle,
serra-garcia2016, % mechanical autonomous stochastic heat engine
gavrilov2016,
rossnagel2016-single-atom-engine,
proesmans2016-prx},
including
the implementation of microscopic single-particle heat engines 
\cite{blickle2012-stochastic-engine,martinez2016-brownian-carnot-engine},
and much theoretical activity
\cite{
sekimoto1998,
sekimoto2000,
hatano-sasa-2001,
speck-seifert2006,
schmiedl2007,
vaikuntanathan-jarzynski2009,
esposito2010-efficiency-carnot,
zimmermann2012-F1-ATPase,
aurell2012, % Refined Second Law of Thermodynamics for Fast Random Processes
horowitz-esposito2014, % Thermodynamics with Continuous Information Flow
zulkowski2014-bit-erasure, % Optimal Finite-Time Erasure of a Classical Bit
verley2014-unlikely-carnot-efficiency,
parrondo2015-thermo-of-info, % Thermodynamics of information
polettini2015-carnot-finite-power, % Efficiency Statistics at All Times: Carnot Limit at Finite Power
altaner2016-fdr-far-from-equilibrium, % Fluctuation-Dissipation Relations Far from Equilibrium
gingrich2016,
gavrilov2017, % Direct measurement of weakly nonequilibrium system entropy is consistent with Gibbs–Shannon form
wulfert2017, % Driven Brownian particle as a paradigm for a nonequilibrium heat bath
gregoire2017, % Stochastic Thermodynamics of Brownian Motion
bonanca2018,
martinez2019, % Inferring broken detailed balance in the absence of observable currents
maes2019, % Nonequilibrium calorimetry
vroylandt2020, % Efficiency Fluctuations of Stochastic Machines Undergoing a Phase Transition
horowitz-gingrich2020,
manikandan2020,
plata2020}.

A fundamental problem in stochastic thermodynamics is to understand how small 
systems do useful work while operating out of equilibrium. A natural framing 
of this problem is in terms of a notion of optimality out of equilibrium, whereby a 
system is considered optimal if it minimizes irreversible heat loss to 
the reservoir on average. Optimal driving protocols can therefore be computed by minimizing
the average dissipation over protocols. 
In general, however, this is a nontrivial optimization problem to solve \cite{seifert2007}.

The introduction of the thermodynamic metric framework \cite{crooks2007,sivak2012} 
simplified the problem 
for a restricted class of systems by recasting it in
a geometric picture in which the average dissipation is proportional to a measure of length in 
the space of control parameters of the system. 
The ``length" is defined by a Riemannian metric on this space. 
An optimal protocol between two points in control 
space is then given by the minimum of this length,
which is generally easier to compute than solutions to the original optimization problem.
This framework is a generalization to mesoscale, out-of-equilibrium systems 
of geometrical approaches originally developed for macroscale, endoreversible systems \cite{
weinhold1975metric,
ruppeiner1979,
burbea1982entropy,
salamon1983thermodynamic,
salamon1984relation,
schlogl1985thermodynamic,
brody1995}.

Since its introduction, the thermodynamic metric framework has found success in predicting
optimal protocols for a number of systems, both analytically and numerically
\cite{zulkowski2012-geometry-thermodynamic-control,
zulkowski2015-optcontrol-overdamped,
rotskoff2015-optcontrol,
sivak2016,
rotskoff2017-geometry-optcontrol-nanomagnet},
and in illuminating
their general characteristics, opening up a window onto the 
physics of small machines that operate out of equilibrium.

The concept of a thermodynamic geometry at mesoscopic length scales emerges 
independently from various different assumptions 
about
the dynamics of the stochastic system. All these approximations
have in common a notion of closeness to equilibrium. 
In the original work, the approximations were linear response plus slow driving~\cite{sivak2012}.
Subsequent work derived a thermodynamic metric under approximations of 
derivative truncation~\cite{zulkowski2012-geometry-thermodynamic-control},
and
timescale separation~\cite{rotskoff2017-geometry-optcontrol-nanomagnet}.
Slow driving was also assumed in order to extend the thermodynamic metric framework to 
driven discrete-time systems \cite{mandal2016}.

In this paper, we provide a rigorous derivation of a thermodynamic metric 
within the framework of the Fokker-Planck
equation for Brownian motion with time-varying control parameters.
We work in a regime in which the control parameters vary on a timescale
that is much longer than the intrinsic timescale of the system, 
which is set by its relaxation time.
The solution to the time-dependent Fokker-Planck equation 
is obtained as an expansion in a small dimensionless parameter $\nu$ that is 
the ratio of the relaxation time of the system to
the shortest characteristic timescale of variation among the 
control parameters. 
The expansion is enabled by the spectral properties of the corresponding
Schr\"{o}dinger operator.
The formula for the thermodynamic metric we derive in this framework
is exact and has a generalization to higher 
dimensions.

In addition, we demonstrate an emergent diffeomorphism symmetry 
in the space of control parameters arising from the
expansion in $\nu$ of the probability density. 
Every term with $n$ indices in the corresponding expansion for the average 
dissipation is a rank $n$ tensor under the diffeomorphism symmetry.

The harmonic potential is a canonical system to 
study in stochastic thermodynamics, both experimentally and 
theoretically~\cite{seifert2007,
imparato2007,
gomez-marin2008,
speck2011,
blickle2012-stochastic-engine,
zulkowski2012-geometry-thermodynamic-control,
zulkowski2013-optcontrol-ness,
kwon2013,
martinez2015-trapped-brownian-particle,
martinez2016-brownian-carnot-engine,
gong2016}.
For this reason we illustrate our formalism and formulas 
using the example of a harmonic oscillator with a time-varying spring constant 
in a time-varying electric field.

\section{Driven Brownian Motion}

Consider a small system 
in contact with a reservoir such as a Brownian 
particle in a suspension subject to an external potential 
$V_{\boldsymbol{\lambda}(t)}(x)$ that can depend on a possibly
time-dependent control vector $\boldsymbol{\lambda}\in\mathbb{R}^k$.
The space $\mathcal{C}$ of all possible values of $\boldsymbol{\lambda}$ 
is a subset of $\mathbb{R}^k$.
The position of the particle 
is given by $x\in\mathbb{R}$ and its probability density $\rho(x; t)$ 
evolves according to a Fokker-Planck equation~\cite{risken1984},
\begin{equation}
    \frac{\partial}{\partial t}\rho(x; t)
    = \hat{\mathcal{L}}_{\boldsymbol{\lambda}(t)}(x)\rho(x; t),
    \label{eq:Fokker-Planck}
\end{equation}
where $\hat{\mathcal{L}}_{\boldsymbol{\lambda}(t)}(x)$, 
the Fokker-Planck operator, is a second-order 
differential operator involving spatial derivatives of the potential. 
In the overdamped limit, where inertial effects are neglected, 
$\hat{\mathcal{L}}_{\boldsymbol{\lambda}(t)}(x)$ takes the form
\begin{equation}
    \hat{\mathcal{L}}_{\boldsymbol{\lambda}(t)}(x)
    = \frac{1}{\gamma}\frac{\partial}{\partial x}
    \bigg(\pot^{\prime}(x)
    + \frac{1}{\beta} \; \frac{\partial}{\partial x}
    \bigg),
    \label{eq:overdamped-FP-operator}
\end{equation}
where $\gamma$ and $\beta = 1/k_BT$ are the friction coefficient and 
inverse temperature, respectively, and $k_B$ is Boltzmann's constant.\footnote{The action of $\FPop$ on $\rho(x;t)$ is 
\begin{equation*}
    \frac{1}{\gamma}\frac{\partial}{\partial x}
    \bigg(\pot^{\prime}(x)\rho(x;t)
    + \frac{1}{\beta} \; \frac{\partial \rho(x;t)}{\partial x}
    \bigg).
\end{equation*}}
Primes denote derivatives with respect to $x$.
Note that $\pot^{\prime}(x)=-F(x;t)$ where $F$ is the force acting on the system.
We consider natural boundary conditions, requiring 
$\rho(x;t)\rightarrow 0$ as $x\rightarrow\pm\infty$.
$\rho(x;t)$ satisfies the normalization condition
\begin{equation}
    \int dx \, \rho(x;t) = 1.
    \label{eq:normalization-of-rho}
\end{equation}
We use the notation $\int dx$ as shorthand for
$\int_{-\infty}^{\infty} dx$ throughout the paper.

Equation\,\eqref{eq:Fokker-Planck} can also be 
written in the form of a continuity equation as 
\begin{equation}
    \frac{\partial}{\partial t}\rho(x; t)=-\frac{\partial}{\partial x}J(x;t),
\end{equation}
where $J$ is the probability current,
\begin{equation}
    J(x;t) = -\frac{1}{\gamma}
    \bigg(\pot^{\prime}(x)
    + \frac{1}{\beta} \; \frac{\partial}{\partial x}
    \bigg)
    \rho(x;t).
    \label{eq:probability-current}
\end{equation}
Natural boundary conditions additionally 
require $J(x;t)\rightarrow 0$ as $x\rightarrow\pm\infty$.

We note that Eq.\,\eqref{eq:Fokker-Planck} with $\FPop$ as given in 
Eq.\,\eqref{eq:overdamped-FP-operator}
is equivalent to the trajectory-level
Langevin description,
\begin{equation}
    \gamma\dot{x}=F(x,t)+\sqrt{\frac{2\gamma}{\beta}}\eta(t),
    \label{eq:langevin-description}
\end{equation}
where 
$\eta(t)$ is mean zero $\delta$-correlated Gaussian noise: 
$\langle\eta(t)\rangle=0$, 
$\langle\eta(t)\eta(t^{\prime})\rangle=\delta(t-t^{\prime})$.
The dot denotes a derivative with respect to time.

At all times, the state space admits the existence of
a unique equilibrium distribution 
$\rhoeq(x)$ such that 
\begin{equation}
    \FPop(x)\rhoeq(x) = 0
    \label{eq:L-on-rhoeq}
\end{equation}
and 
\begin{equation}
    \int dx \, \rhoeq(x) = 1.
    \label{eq:normalization-of-rhoeq}
\end{equation}
$\rhoeq(x)$ is given by
\begin{equation}
    \rhoeq(x) = \frac{1}{Z(t)}e^{-\beta\pot(x)},
    \label{eq:rhoeq}
\end{equation}
where $Z(t)$ is the partition function,
\begin{equation}
    Z(t)=\int dx \, e^{-\beta\pot(x)}.
\end{equation}
All distributions approach $\rhoeq(x)$ 
asymptotically with time when $\boldsymbol{\lambda}$ is frozen, and
$\rhoeq$ satisfies the detailed balance condition, which requires that the probability current
in equilibrium be zero,
\begin{equation}
    -\frac{1}{\gamma}
    \bigg(\pot^{\prime}(x)
    + \frac{1}{\beta} \; \frac{\partial}{\partial x}
    \bigg)
    \rhoeq(x)
    =0 \,\, \forall x.
\end{equation}

We note that $\rhoeq$ does not satisfy Eq.\,\eqref{eq:Fokker-Planck} except
in an approximate sense.
While Eq.\,\eqref{eq:L-on-rhoeq} is exact, 
the time derivative of $\rhoeq$ is
\begin{equation}
    \frac{\partial}{\partial t}\rhoeq(x)
    =\sum_{i=1}^{k}\dot{\lambda}_i\frac{\partial}{\partial\lambda_i}\rhoeq(x),
    \label{eq:time-deriv-of-rhoeq}
\end{equation}
which is not zero if $\dot{\lambda}_i\neq 0$.
The solution to Eq.\,\eqref{eq:Fokker-Planck} that we develop in the following
is in the limit of small $\dot{\boldsymbol{\lambda}}$.
We will show that the ``smallness" of $\dot{\boldsymbol{\lambda}}$ is quantified by a
parameter $\nu$, defined as the ratio of 
the relaxation time $\tau_{\alpha_1}$ of the system to the driving 
timescale $\tau_{\lambda}$, which must be chosen such that $\nu\ll 1$.
In this limit, the timescale of driving is so long that $\rhoeq$ is roughly 
stationary on the system timescale, which is set by $\tau_{\alpha_1}$.
Thus, $\rhoeq$ satisfies Eq.\,\eqref{eq:Fokker-Planck} to zeroth order 
in the parameter $\nu$.
We return in detail to these ideas in Sec.\,\ref{sec:expansion-parameter}.

We solve Eq.\,\eqref{eq:Fokker-Planck} using the method of Green's functions. The difficulty in this
program is that the Fokker-Planck operator has a zero mode, namely, $\rhoeq$, 
and is not self-adjoint.
We map $\FPop$ onto its corresponding Schr\"{o}dinger operator, which is self-adjoint,
and leverage the spectral theory of the latter to construct the Green's function of $\FPop$.

For the purposes of solving Eq.\,\eqref{eq:Fokker-Planck}, the partial
derivative with respect to time on the left-hand side
should be interpreted as acting at fixed $\boldsymbol{\lambda}$.
We will show in Sec.\,\ref{sec:expansion-parameter}
that this produces a solution that is \textit{consistent}, in the sense that both the
left-hand side of Eq.\,\eqref{eq:Fokker-Planck} and the time derivative
of the solution we find to this equation are $\mathcal{O}(\nu)$.

\subsection{The associated Schr\"{o}dinger operator and Green's function}

The Fokker-Planck operator $\FPop$ is not self-adjoint. However, we can construct a 
self-adjoint operator $\Sym$ from $\FPop$ by making the similarity
transformation
\begin{equation}
    \Sym 
    = e^{\beta \pot/2}\FPop e^{-\beta \pot/2}.
    \label{eq:symmetrizing-transform-FP}
\end{equation}
We have suppressed the $x$ dependence of the potential and the operators for notational
convenience.
$\Sym$ and $\FPop$ share eigenvalues, and their eigenfunctions are
related by a simple transformation that we will discuss shortly.
$\Sym$ takes the form
\begin{equation}
    \Sym(x) = \frac{1}{\gamma\beta}\bigg(
    \frac{\beta}{2}\pot^{\prime\prime}(x) 
    - \left(\frac{\beta}{2}\pot^{\prime}(x)\right)^2
    + \frac{\partial^2}{\partial x^2}\bigg).
    \label{eq:symmetric-operator}
\end{equation}
It is related to the one-dimensional single-particle Schr\"{o}dinger operator 
$\Schr$ as follows:
\begin{equation}
    \Schr = -\frac{1}{2}\Sym.
    \label{eq:schrodinger-to-H}
\end{equation}
We have
\begin{equation}
    \Schr = -\frac{1}{2\gamma\beta}\frac{\partial^2}{\partial x^2} + \Schrpot(x),
    \label{eq:schrodinger}
\end{equation}
where the potential $\Schrpot$ is given by
\begin{equation}
    \Schrpot(x) = \frac{1}{2\gamma\beta} \left(
    \left(\frac{\beta}{2} \pot^{\prime}(x)\right)^2 -\frac{\beta}{2}\pot^{\prime\prime}(x) \right).
    \label{eq:Schrpot}
\end{equation}

The map we have described between Fokker-Planck operators and Schr\"{o}dinger
operators is well-known \cite{risken1971,pavliotis2014}.
We use it here to apply the spectral theory of the Schr\"{o}dinger
operator to driven Brownian motion. Any potential for which the 
spectral decomposition
of the Schr\"{o}dinger operator is known and possesses certain properties
then becomes accessible 
to us for the purposes of solving Eq.\,\eqref{eq:Fokker-Planck}.

As mentioned, the requirements for this approach to be viable involve 
conditions on the 
spectrum of $\Schr$.
Natural boundary conditions on Eq.\,\eqref{eq:Fokker-Planck} already require 
$\pot(x)\rightarrow\infty$ as $x\rightarrow\pm\infty$.
We additionally require $\pot$ to be such that $\Schrpot$ is also confining. That is,
$\Schrpot(x)\rightarrow\infty$ as $x\rightarrow\pm\infty$. 
This is satisfied, for example, if $\pot$ is harmonic,
and not satisfied if it is logarithmic in $|x|$ at large $x$.

We use $E_n$ and $\psi_n$ to denote the eigenvalues and eigenfunctions of $\Schr$.
The eigenvalue equation is
\begin{equation}
    \Schr(x)\psi_n(x) = E_n\psi_n(x), \, n=0,1,\dots.
    \label{eq:eigensystem-schr}
\end{equation}
For $x\in\mathbb{R}$, with the stated boundary condition on $\Schrpot$, we are
guaranteed that the spectrum of $\Schr$ is discrete, nondegenerate 
($E_m \neq E_n$ for $m\neq n$), and ordered ($E_n < E_{n+1} \, \forall n$).
The fact that a confining potential confers a discrete nondegenerate spectrum can be proved rigorously (see Theorem 10.7 in \cite{hislop-sigal}).
From a physical point of view this is reasonable to expect because in one spatial dimension a
confining potential has bounded closed orbits which are quantized to give 
a discrete nondegenerate spectrum. (Tunneling effects can split degenerate energy levels separated by a potential barrier.)
The discreteness of the spectrum crucially enables a simple definition of the
Green's function of $\Schr$.
See \cite{landau-lifschitz-nonrelativistic-QM} for a proof of nondegeneracy.

It is simple to check\footnote{Schr\"{o}dinger operators customarily have nonzero zero-point energies.
Here, $E_0=0$ due to the specific construction of $\Schrpot$, which is ``shifted"
downward by a factor of $\pot^{\prime\prime}/4\gamma$ such that the
usual zero-point energy of Eq.\,\eqref{eq:schrodinger} is exactly removed.}
that $E_0=0$ and 
that the zeroth eigenfunction of $\Schr$ is given by
\begin{equation}
    \psi_0(x) = \frac{1}{\sqrt{Z(t)}}e^{-\beta\pot(x)/2}.
\end{equation}
Note that $\rhoeq = \psi_0^2$.
The $\psi_n$ are real and form a complete orthonormal basis 
\cite{landau-lifschitz-nonrelativistic-QM}:
\begin{equation}
    \int dx\; \psi_n(x)\psi_m(x)=\delta_{nm},
    \label{eq:orthogonality-of-psi}
\end{equation}
where $\delta_{nm}$ is the Kronecker delta.
This guarantees the representation 
\begin{equation}
    \delta(x-y)=\sum_n \psi_n(x)\psi_n(y)
    \label{eq:completeness-of-psi}
\end{equation}
for the delta function.

For $n>0$, the eigenvalues of $\Schr$ satisfy $E_{n} > 0$.
The proof of this claim is as follows. By left-multiplying Eq.\,\eqref{eq:eigensystem-schr}
by $\psi_n$ and integrating with respect to $x$,
we have
\begin{equation}
    E_n = \int dx  \, \left(\frac{1}{2\gamma\beta}\left(\frac{\partial \psi_n}{\partial x}\right)^2+
    \Schrpot\psi_n^2\right).
\end{equation}
Writing $\psi_n(x) = \rho_{l,n}(x)\psi_0(x)$ where $\rho_{l,n}$ is a smooth function with $n$ nodes,
this is
\begin{equation}
    E_n = \int dx \, \frac{1}{2\gamma\beta}
    \psi_0^2\left(\frac{\partial \rho_{l,n}}{\partial x}\right)^2 \geq 0,
\end{equation}
with equality holding only for $n=0$ since $\rho_{l,0}=1$.
The subscript $l$ notation will become clear in the next section.

The function $\rho_{l,n}$ satisfies the eigenvalue equation
\begin{equation}
    \frac{1}{\gamma}\left(-\pot^{\prime}(x)+
    \frac{1}{\beta}\frac{\partial}{\partial x}\right)\frac{\partial \rho_{l,n}}{\partial x}
    = \adjointFPop \rho_{l,n}=-2E_n \rho_{l,n},
    \label{eq:backward-Kolmogorov}
\end{equation}
where $\adjointFPop$ is
the Kolmogorov backward operator.\footnote{This operator is self-adjoint under the measure $dm(x)$ defined by $dm(x)=\left(\rhoeq(x)\right)^{-1}dx$.} 
$\adjointFPop$ satisfies the symmetrization relation
\begin{equation}
    \Sym 
    = e^{-\beta \pot/2}\adjointFPop e^{\beta \pot/2}.
    \label{eq:symmetrizing-transform-adjoint}
\end{equation}

Given the structure of the spectrum of $\Schr$, its Green's function $\SchrGreen(x;y)$ 
is given by the following standard definition:
\begin{equation}
    \SchrGreen(x;y) = \sum_{n\neq 0}\frac{1}{E_n}\psi_n(x)\psi_n(y).
    \label{eq:greens-fn-of-schr}
\end{equation}
The action of $\Schr$ on $\SchrGreen$ is
\begin{equation}
    \Schr(x)\SchrGreen(x;y) = \delta(x-y) - \psi_0(x)\psi_0(y).
    \label{eq:greens-fn-of-Schr}
\end{equation}
Note that the right-hand side of Eq.\,\eqref{eq:greens-fn-of-schr} has the form of a 
projection. It indicates that $\Schr$ is only invertible in the space of functions
orthogonal to $\psi_0$.

$\Sym$ and $\Schr$ share eigenfunctions $\psi_n$. Writing $\alpha_n$ for the eigenvalues of
$\Sym$, these are given by
\begin{equation}
    \alpha_n = -2E_n,
    \label{eq:alpha-and-E}
\end{equation}
where $\alpha_0=0$ and $\alpha_{n>0}<0$. The eigenvalue equation 
for $\Sym$ is
\begin{equation}
    \Sym(x)\psi_n(x) = \alpha_n\psi_n(x).
    \label{eq:eigensystem-H}
\end{equation}
The Green's function $G_{\mathcal{H}}$ of $\Sym$ is given by Eq.\,\eqref{eq:greens-fn-of-schr}
with the replacement $E_n\rightarrow\alpha_n$:
\begin{equation}
    G_{\mathcal{H}}(x;y)=\sum_{n\neq 0}\frac{1}{\alpha_n}\psi_n(x)\psi_n(y).
    \label{eq:greens-fn-of-H}
\end{equation}
The action of $\Sym$ on $G_{\mathcal{H}}$ is
\begin{equation}
    \Sym(x)G_{\mathcal{H}}(x;y) = \delta(x-y) - \psi_0(x)\psi_0(y).
    \label{eq:action-of-GH}
\end{equation}

\subsection{The Green's function of $\FPop$}

We use the discussion of the previous section
to write down the eigenfunctions of $\FPop$ and $\adjointFPop$,
and the Green's function of $\FPop$.

From Eqs.\,\eqref{eq:symmetrizing-transform-FP}, 
\eqref{eq:symmetrizing-transform-adjoint},
and \eqref{eq:eigensystem-H},
we immediately have the relations
\begin{subequations}
\begin{align}
    \hat{\mathcal{L}}(x)\rho_{r, n}(x) 
    &= \alpha_n\rho_{r, n}(x), \label{eq:right-rho}\\
    \hat{\mathcal{L}}^{\dagger}(x)\rho_{l, n}(x) 
    &= \alpha_n\rho_{l, n}(x),\label{eq:left-rho}
\end{align}
\end{subequations}
where
\begin{subequations}\label{eq:defn-right-and-left-eigenfunctions}
\begin{align}
    \rho_{r, n}(x) &= \psi_0(x)\psi_n(x),\\
    \rho_{l, n}(x) &= \left(\psi_0(x)\right)^{-1}\psi_n(x).
\end{align}
\end{subequations}
$\rho_{r, n}$ and $\rho_{l, n}$ are called the right and left eigenfunctions, respectively.
Together, they form a biorthogonal system that diagonalizes $\FPop$.
They are complete, 
\begin{equation}
    \delta(x-y) = \sum_n \rho_{r, n}(x)\rho_{l, n}(y),
    \label{eq:completeness-of-rho}
\end{equation}
and orthonormal,
\begin{equation}
    \int dx \; \rho_{r, n}(x)\rho_{l, m}(x) = \delta_{nm}.
    \label{eq:orthogonality-of-rho}
\end{equation}
Equation\,\eqref{eq:completeness-of-rho} follows from Eq.\,\eqref{eq:completeness-of-psi},
and Eq.\,\eqref{eq:orthogonality-of-rho} follows from Eqs.\,\eqref{eq:orthogonality-of-psi}
and \eqref{eq:completeness-of-rho}.
The zeroth right eigenfunction is the equilibrium distribution 
of $\FPop$ corresponding to the specific value of $\boldsymbol{\lambda}$ 
at time $t$, and the zeroth left eigenfunction is a constant:
\begin{equation}
    \rho_{r,0}(x) = \psi_0^2(x)=\rhoeq(x), \, \, \rho_{l,0}(x) = 1.
    \label{eq:zeroth-eigenfunctions}
\end{equation}
Due to these last two facts, the right and left eigenfunctions share the
simple relationship
\begin{equation}
    \rho_{r, n} = \rho_{r, 0}\rho_{l, n}.
    \label{eq:left-rho-trick}
\end{equation}

We can now write the Green's function $\Green$ of $\FPop$.
Using the representation given by Eq.\,\eqref{eq:symmetrizing-transform-FP} for $\Sym$,
and suppressing the subscript $\boldsymbol{\lambda}(t)$ for visual clarity,
from Eq.\,\eqref{eq:action-of-GH} we have
\begin{equation}
    e^{\beta V(x)/2}\FPopshort(x) e^{-\beta V(x)/2} G_{\mathcal{H}}(x;y) = \sum_{n\neq 0}\psi_n(x)\psi_n(y).
\end{equation}
By left-multiplying by $e^{-\beta V(x)/2}$, right-multiplying by $e^{\beta V(y)/2}$,
and using Eq.\,\eqref{eq:defn-right-and-left-eigenfunctions}, we arrive at
\begin{equation}
    \FPopshort(x) e^{-\beta V(x)/2} G_\mathcal{H}(x;y) e^{-\beta V(y)/2} = \sum_{n\neq 0}\rho_{r,n}(x)\rho_{l,n}(y),
    \label{eq:defn-of-G-intermediate}
\end{equation}
from which we identify $\Green$:
\begin{align}
    \Green(x;y) &= e^{-\beta \pot(x)/2} G_{\mathcal{H}}(x;y) e^{\beta \pot(y)/2} \nonumber\\
    &=\sum_{n\neq 0} \frac{1}{\alpha_n}\rho_{r,n}(x)\rho_{l,n}(y).
    \label{eq:defn-of-G}
\end{align}
The action of $\FPop$ on $\Green$ is given by Eq.\,\eqref{eq:defn-of-G-intermediate}.
Using Eqs.\,\eqref{eq:completeness-of-rho} and \eqref{eq:zeroth-eigenfunctions}, 
this can be rewritten as
\begin{equation}
    \FPop(x)\Green(x;y)=\delta(x-y)-\rhoeq(x).
    \label{eq:action-of-L-on-G}
\end{equation}

\subsection{Solution to the Fokker-Planck equation}

We can decompose the probability distribution in Eq.\,\eqref{eq:Fokker-Planck}
into the sum of $\rhoeq(x)$ and a correction $\delta\rho(x; t)$,
\begin{equation}
    \rho(x; t) = \rho^{eq}_{\boldsymbol{\lambda}(t)}(x) + \delta\rho(x; t).
    \label{eq:decomposition-of-rho}
    \end{equation}
We must have $\int dx \; \delta\rho(x; t)=0$ to preserve normalization. 
Using this representation for $\rho(x;t)$ in Eq.\,\eqref{eq:Fokker-Planck},
we obtain the dynamics of $\delta\rho(x;t)$,
\begin{equation}
    \FPop(x)\delta\rho(x; t) = \frac{\partial}{\partial t}\rho(x;t).
    \label{eq:dynamics-deltarho}
\end{equation}
In order to apply the method of Green's functions, we interpret the right-hand side of
Eq.\,\eqref{eq:dynamics-deltarho} as a source term.  
From this follows the solution
\begin{equation}
    \delta\rho(x;t)
    =\int dy \; 
    \Green(x;y)
    \frac{\partial}{\partial t}\rho(y; t).
    \label{eq:solution-to-delta-rho}
\end{equation}
Equation\,\eqref{eq:solution-to-delta-rho} contains the quantity 
$\delta\rho$ on both sides and can be solved iteratively. 
Thus we 
arrive at the solution
\begin{widetext}
\begin{align}
    \rho(x; t)
    &=\rho^{eq}_{\boldsymbol{\lambda}(t)}(x)
    +\int dx^{\prime} \; 
        \Green(x;x^{\prime})
        \frac{\partial}{\partial t}
        \rho^{eq}_{\boldsymbol{\lambda}(t)}(x^{\prime})
    +\int dx^{\prime\prime} \; 
        \Green(x;x^{\prime\prime}) \; 
        \frac{\partial}{\partial t}\int dx^{\prime} \;
        \Green(x^{\prime\prime};x^{\prime})
        \frac{\partial}{\partial t}
        \rho^{eq}_{\boldsymbol{\lambda}(t)}(x^{\prime})
    +\dots,
    \label{eq:solution-to-FP-equation}
\end{align}
\end{widetext}
with the partial time derivative of $\rhoeq$ given by 
Eq.\,\eqref{eq:time-deriv-of-rhoeq}.

The form of Eq.\,\eqref{eq:solution-to-FP-equation} is 
$\rho(x;t)= \rhoeq(x)+\sum_{n=1}^{\infty}\delta\rho^{(n)}(x;t)$, 
where the quantities $\delta\rho^{(n)}$ are corrections to $\rhoeq$.
We observe that the corrections have a 
recursive structure, and integrate to zero:
\begin{subequations}
\begin{align}
    &\delta\rho^{(n+1)}(x; t)
    =\int dx^{\prime} \; 
        \Green(x;x^{\prime}) \; 
        \frac{\partial}{\partial t}
        \delta\rho^{(n)}(x^{\prime}; t), 
    \label{eq:recursive-structure-of-deltarho}\\
    &\int dx \; \delta\rho^{(n+1)}(x; t)=0, \, \, n\geq0.
    \label{eq:deltarho-integrates-to-zero}
\end{align}
\end{subequations}
In the above, we have notated $\rhoeq(x)$ as $\delta\rho^{(0)}(x;t)$.
The form of Eq.\eqref{eq:recursive-structure-of-deltarho}
indicates that $\delta\rho^{(n+1)}(x; t)$ contains precisely $n+1$ derivatives
with respect to time. This motif will be important in Sec.\,\ref{sec:derivation-of-metric},
where we will see that it introduces geometric structure
to the average dissipation.

\subsection{The expansion parameter $\nu$}
\label{sec:expansion-parameter}

Equation\,\eqref{eq:solution-to-FP-equation} is a derivative
expansion.
In this section, we justify this claim.

There are two sources of timescales in this problem: the eigenvalues of the
Fokker-Planck operator, and the time variation of the control parameters.

The eigenvalues $\alpha_n$ of $\FPop$ have the
physical units of inverse time, and their 
absolute values set the various natural timescales of the system.
Calling these timescales $\tau_{\alpha_n}$, we have $\tau_{\alpha_{n}}=1/|\alpha_{n}|$.
Due to the ordering of the $\alpha_n$, the $\tau_{\alpha_n}$ are also ordered.
The longest natural timescale in the system is $\tau_{\alpha_1}$,
known as the relaxation time.

Each external parameter $\lambda_i$ has a characteristic timescale 
$\tau_{\lambda_i}$ associated with its time evolution.
We denote the shortest of these timescales as
$\tau_{\lambda}=\min_i \tau_{\lambda_i}$.

Now let us examine the total time variation of $\rho(x;t)$,
\begin{equation}
    \frac{d}{dt}\rho(x;t)
    =\frac{\partial}{\partial t}\rho(x;t) 
    + \sum_i \dot{\lambda}_i\frac{\partial}{\partial \lambda_i}
    \rho(x;t). 
    \label{eq:total-time-deriv-rho}
\end{equation}
In the first term on the right-hand side of
Eq.\,\eqref{eq:total-time-deriv-rho}, the time derivative acts at 
fixed $\boldsymbol{\lambda}$ and the time evolution is generated by 
the Fokker-Planck operator, i.e., by
Eq.\,\eqref{eq:Fokker-Planck}. 
The second term describes the time variation resulting from the 
time dependence of the 
external control parameters, which is not determined by the Fokker-Planck operator.\footnote{We will see in a later section that this time variation is determined by another principle, namely, the minimization of the average heat produced in the reservoir over the course of driving.}
Note that if we replace $\rho(x;t)$ by $\rhoeq(x)$ in Eq.\,\eqref{eq:total-time-deriv-rho},
the first term on the right-hand side evaluates to zero, exactly consistent
with Eq.\,\eqref{eq:time-deriv-of-rhoeq}.

In this work, we consider the scenario in which the dynamics of 
$\boldsymbol{\lambda}$ is very slow compared to the dynamics generated 
by the Fokker-Planck operator.
This means the longest natural timescale $\tau_{\alpha_1}$ 
must be shorter than the shortest control timescale $\tau_{\lambda}$:
\begin{equation}
    \tau_{\lambda}\gg\tau_{\alpha_1}.
    \label{eq:slow-variation-constraint}
\end{equation}
Equation\,\eqref{eq:slow-variation-constraint} naturally gives rise to a 
dimensionless small parameter $\nu$, defined as follows:
$\nu=\tau_{\alpha_1}/\tau_{\lambda}\ll 1$.
It is the smallness of this parameter that justifies our usage of 
Eq.\,(1) to approximate the true dynamics of $\rho(x;t)$, which
is given by the left-hand side of Eq.\,\eqref{eq:total-time-deriv-rho}.

In Eq.\,\eqref{eq:solution-to-FP-equation}, derivatives with respect to time act 
(through $\Green$ and $\rhoeq$) only on $\boldsymbol{\lambda}(t)$, and so
we can rescale time in $\boldsymbol{\lambda}-$space by $\nu$ by defining
the variable $\Tilde{t}=\nu t$. Making the reparametrization $t\rightarrow\Tilde{t}$
in Eq.\,\eqref{eq:solution-to-FP-equation},
we arrive at an expansion for $\rho(x;t)$ in the manifestly dimensionless 
small parameter $\nu$:
\begin{widetext}
\begin{align}
    \rho(x; \Tilde{t})
    &=\rho^{eq}_{\boldsymbol{\lambda}(\Tilde{t})}(x)
    +\nu\int dx^{\prime} \; 
        G_{\boldsymbol{\lambda}(\Tilde{t})}(x;x^{\prime})
        \frac{\partial \rho^{eq}_{\boldsymbol{\lambda}(\Tilde{t})}}{\partial \Tilde{t}}
        (x^{\prime})
    +\nu^2\int dx^{\prime\prime} \; 
        G_{\boldsymbol{\lambda}(\Tilde{t})}(x;x^{\prime\prime}) \; 
        \frac{\partial}{\partial \Tilde{t}}\int dx^{\prime} \;
        G_{\boldsymbol{\lambda}(\Tilde{t})}(x^{\prime\prime};x^{\prime})
        \frac{\partial \rho^{eq}_{\boldsymbol{\lambda}(\Tilde{t})}}{\partial \Tilde{t}}(x^{\prime})
    +\dots.
    \label{eq:final-expansion-for-rho}
\end{align}
\end{widetext}

What is happening here is that there is a separation of 
timescales between the laboratory and the control space. In the latter, 
time must be measured in units of $\tau_{\lambda}$. However, the overall
timescale of the problem is set by $\tau_{\alpha_1}$, which is fixed
by the shape of the potential. Therefore when expanding the density
$\rho(x;t)$, it is necessary to measure $\tau_{\lambda}$ 
\emph{in units of} $\tau_{\alpha_1}$. This is why time in control
space is scaled by $\nu$.

The condition given by Eq.\,\eqref{eq:slow-variation-constraint} imposes
a constraint on the dynamics of the spectrum of $\FPop$, which we now discuss.
In general, the $\alpha_n$ are functions of all the control parameters $\lambda_i$
due to the fact that the spectrum of $\FPop$ depends on $\pot$, which is a function
of $\boldsymbol{\lambda}$.
The time derivative of $\alpha_n$ is
\begin{equation}
    \frac{d \alpha_n}{dt} = \sum_{i}\dot{\lambda}_i\frac{\partial \alpha_n}{\partial \lambda_i},
    \label{eq:time-deriv-of-En}
\end{equation}
where the variation of $\alpha_n$ with respect to $\lambda_i$ is given by the 
Hellmann-Feynman theorem \cite{feynman1939}
\begin{equation}
    \frac{\partial \alpha_n}{\partial \lambda_i} 
    = \int dx \; \psi^2_n(x) \frac{\partial \Sym(x)}{\partial \lambda_i} 
    = -2\int dx \; \psi^2_n(x) \frac{\partial \Schrpot(x)}{\partial \lambda_i}.
    \label{eq:hellmann-feynman}
\end{equation}
For every $i\in(1,\dots,k)$,
Eq.\,\eqref{eq:hellmann-feynman} is finite and fully 
determined by the form of the potential $\Schrpot$. Therefore
Eq.\,\eqref{eq:slow-variation-constraint}, which can equivalently be written as
$\max_{i}\left|\dot{\lambda}_i\right|\ll |\alpha_1|$, 
together with Eq.\,\eqref{eq:time-deriv-of-En}, implies that the quantities
$\left|\dot{\alpha}_n\right|$
must be small $\forall n$.
We can explicitly check that this condition holds. Note that
      \begin{equation}
    \dot{\lambda}_i
    =\frac{d\Tilde{t}}{dt}\frac{d\lambda}{d\Tilde{t}}
    =\nu\frac{d\lambda}{d\Tilde{t}}
    =\mathcal{O}(\nu),
    \label{eq:lambda-dot}
\end{equation}
and so $\dot{\lambda}_i$ is of the order of $\nu$.
Together with Eq.\,\eqref{eq:lambda-dot}, Eq.\,\eqref{eq:time-deriv-of-En}
implies that $\left|\dot{\alpha}_n\right|$ is also $\mathcal{O}(\nu)$.
That is, the condition given by Eq.\,\eqref{eq:slow-variation-constraint} forces 
the spectrum of $\FPop$ to change slowly over the course of driving.

Due to the fact that derivatives with respect
to time in Eq.\,\eqref{eq:final-expansion-for-rho} act only on $\boldsymbol{\lambda}(t)$,
Eq.\,\eqref{eq:lambda-dot} also implies that the time derivative of 
Eq.\,\eqref{eq:final-expansion-for-rho} is $\mathcal{O}(\nu)$,
which is consistent with the time dependence of 
Eq.\,\eqref{eq:total-time-deriv-rho} on $\boldsymbol{\lambda}$.

The last point we must address in this timescale analysis is the fact
that $\nu$ itself is a function of time. 
Clearly, in order for the expansion in 
Eq.\,\eqref{eq:final-expansion-for-rho}
to be stable, we require the time variation of $\nu$ to be small.
We can check that Eq.\,\eqref{eq:slow-variation-constraint} indeed
enforces this. Using Eq.\,\eqref{eq:lambda-dot}, we find that
\begin{equation}
    \frac{d\nu}{dt}
    =\mathcal{O}\left(\nu^2\right).
\end{equation}
In fact, the $n^{th}$ time derivative
of $\nu$ for $n\geq 1$ is of the order of $\nu^{n+1}$.

Thus as long as the control timescale is chosen such that the slowness
condition given by Eq.\,\eqref{eq:slow-variation-constraint} is satisfied, the
procedure we have presented for solving Eq.\,\eqref{eq:Fokker-Planck}
is consistent, and Eq.\,\eqref{eq:final-expansion-for-rho}
describes the time evolution of $\rho(x;t)$.

In the next section, we derive a formula for the thermodynamic metric using 
Eq.\,\eqref{eq:solution-to-FP-equation}. We note that in
all previous work 
\cite{sivak2012,zulkowski2012-geometry-thermodynamic-control,rotskoff2017-geometry-optcontrol-nanomagnet}
in which the thermodynamic metric has been derived it is assumed
that the timescale of driving is slow with respect to the longest natural timescale of the
system. The analysis just given explains why this assumption is necessary:
without it, the Fokker-Planck equation is not a good descriptor of the driven Brownian system.

Lastly, we note that other authors have previously made use of eigenfunction expansions 
of $\rho(x;t)$ to calculate the average dissipation
for driven Brownian systems with a single slowly varying control parameter
\cite{sekimoto-sasa-1997,koide2016}.
We will calculate the average dissipation in the next section.
The authors recognized that their methods must correspond to a perturbative
approach to solving Eq.\,\eqref{eq:Fokker-Planck} as we have presented here, 
but this idea was not fully developed. In particular, 
the precise conditions under which the spectral structure 
of $\FPop$ permits a perturbative expansion of $\rho(x;t)$ in $\nu$ and 
the relative importance of the various timescales in the problem were not studied,
and $\tau_{\lambda}$ was not identified.

\section{The thermodynamic metric}

Writing a driving protocol for a system involves 
specifying a functional form for the time dependence of the control 
vector $\boldsymbol{\lambda}$. 
We say a driving protocol $\boldsymbol{\Lambda}$ is optimal if 
it minimizes the functional for the average heat 
$\langle \Delta Q\rangle\left[\boldsymbol{\Lambda}\right]$ produced in the reservoir
over the course of driving \cite{sivak2012},
\begin{equation}
    \boldsymbol{\Lambda}^{\text{opt}}=\argmin_{\boldsymbol{\Lambda}} \,
    \langle \Delta Q\rangle\left[\boldsymbol{\Lambda}\right].
    \label{eq:defn-of-lambda-opt}
\end{equation}
We are interested in the scenario where the system is driven
between two fixed values of $\boldsymbol{\lambda}$ over a fixed time period $\Omega$.
Note that we must have $\Omega \gg \tau_{\lambda}$.

The average heat transferred to the 
reservoir 
over the course of driving
is given by the formula~\cite{sekimoto1997}
\begin{align}
    \langle \Delta Q\rangle \left[\boldsymbol{\Lambda}\right]
    &=-
    \int_0^{\Omega} dt\int dx \,
    \pot^{\prime}(x)J(x;t) \nonumber\\
    &=
    \int_0^{\Omega} dt\int dx \,
    \rho(x; t)
    \left(\frac{\pot^{\prime 2}(x)}{\gamma}
    - 
    \frac{\pot^{\prime\prime}(x)}{\gamma\beta}\right).
    \label{eq:heat-production}
\end{align}
In the second equality, we have replaced $J(x;t)$ with the right-hand side of
Eq.\,\eqref{eq:probability-current} and integrated by parts.
Note that the quantity in parentheses in Eq.\,\eqref{eq:heat-production} 
is, up to a constant factor $4/\beta$, the Schr\"{o}dinger potential $\Schrpot$ at 
inverse temperature $2\beta$.

In the following, 
we calculate $\langle \Delta Q\rangle$ using the approximation
\begin{equation}
    \rho(x; t)
    =\rho^{eq}_{\boldsymbol{\lambda}(t)}(x) 
    +\delta\rho^{(1)}(x; t)
    +\delta\rho^{(2)}(x; t),
    \label{eq:rho-approximation}
\end{equation}
with the corrections $\delta\rho^{(1)}(x; t)$ and $\delta\rho^{(2)}(x; t)$ 
given by the second and third terms on the right-hand side of
Eq.\,\eqref{eq:solution-to-FP-equation}, respectively,
\begin{subequations}
\begin{align}
    &\delta\rho^{(1)}(x; t) 
    = \int dx^{\prime} \; 
        \Green(x;x^{\prime})
        \frac{\partial}{\partial t}
        \rho^{eq}_{\boldsymbol{\lambda}(t)}(x^{\prime}),
    \label{eq:deltarho1}
    \\
    &\delta\rho^{(2)}(x; t)
    =\int dx^{\prime} \; 
        \Green(x;x^{\prime}) \; 
        \frac{\partial}{\partial t}
        \delta\rho^{(1)}(x^{\prime}; t).
    \label{eq:deltarho2}
\end{align}
\end{subequations}
We show that one of the contributions to 
$\langle \Delta Q\rangle$ coming from $\delta\rho^{(2)}$
contains an integral over a symmetric positive definite matrix in the space 
of control parameters $\mathcal{C}$, and we identify this as the thermodynamic
metric for systems described by Eq.\,\eqref{eq:Fokker-Planck} with the stated
conditions on $\pot$ and $\Schrpot$.
We discuss the emergence of this geometric structure in 
$\langle \Delta Q\rangle$ and show that it persists to all orders in the
expansion 
of $\rho(x;t)$
(Eq.\,\eqref{eq:solution-to-FP-equation}).

\subsection{Calculation of $\left<\Delta Q\right>$ and derivation of thermodynamic metric}
\label{sec:derivation-of-metric}

We drop the subscript $\boldsymbol{\lambda}(t)$
for visual clarity.

It is useful to rewrite
Eq.\,\eqref{eq:heat-production} in the equivalent form
\begin{equation}
    \left<\Delta Q\right>[\boldsymbol{\Lambda}] 
    = \frac{1}{\gamma\beta^2}
    \int_0^{\Omega} dt 
    \int dx \; \rho(x; t)
    e^{\beta V(x)}
    \partial_x^2
    e^{-\beta V(x)}.
    \label{eq:heat-production-exp-form}
\end{equation}
The first contribution to $\langle\Delta Q\rangle$ from 
Eq.\,\eqref{eq:rho-approximation} corresponds to approximating $\rho(x;t)$
by $\rhoeqshort(x)$, and it evaluates to zero,
\begin{equation}
    \left<\Delta Q\right>_0
    = \frac{1}{\gamma\beta^2}
    \int_0^{\Omega} dt \int dx \; 
    \rhoeqshort(x)\;
    e^{\beta V(x)}
    \partial^2_{x}e^{-\beta V(x)}
    =0.
    \label{eq:q0}
\end{equation}
This is easily seen by using Eq.\,\eqref{eq:rhoeq} to replace $e^{-\beta V(x)}$ and applying
the normalization condition given by Eq.\,\eqref{eq:normalization-of-rhoeq}.

To calculate the next two terms of $\left<\Delta Q\right>$, we will make use
of the following identity:
\begin{align}
    \int dx \; &G(x;x^{\prime})e^{\beta V(x)}
    \partial^2_{x}e^{-\beta V(x)}\nonumber\\
    &=\gamma\beta \int dx \; \left(1-\beta V(x)\right)
    \FPopshort(x)G(x;x^{\prime}).
    \label{eq:trick1}
\end{align}
This is derived by integrating the left-hand side
by parts twice, evaluating
the resulting double derivative over the product $G(x;x^{\prime})e^{\beta V(x)}$,
and integrating by parts again.
The boundary terms in Eq.\,\eqref{eq:trick1} vanish.

The second contribution to $\left<\Delta Q\right>$ is
\begin{equation}
    \left<\Delta Q\right>_1 
    = \frac{1}{\gamma\beta^2}\int_0^{\Omega} dt 
    \int dx \;
    \delta\rho^{(1)}(x;t)\;
    e^{\beta V(x)}
    \partial^2_{x}e^{-\beta V(x)}.
\end{equation}
Replacing $\delta\rho^{(1)}$ with Eq.\,\eqref{eq:deltarho1}, applying Eq.\,\eqref{eq:trick1} and then Eq.\,\eqref{eq:action-of-L-on-G}, we have
\begin{align}
    \left<\Delta Q\right>_1 
    &= \frac{1}{\beta} \int_0^{\Omega} dt  \int dx^{\prime} \;
    \partial_t\rhoeqshort(x^{\prime})
    \int dx \; \nonumber\\
    &~~~\left(1-\beta V(x)\right)
    \FPopshort(x)G(x;x^{\prime})
    \nonumber\\
    &=
    - \frac{1}{\beta}\int_0^{\Omega} dt \int dx^{\prime} \; 
    \partial_t\rhoeqshort(x^{\prime})\beta V(x^{\prime})\nonumber\\
    &~~~+ \frac{1}{\beta}\int_0^{\Omega} dt  \int dx^{\prime} \;
    \partial_t\rhoeqshort(x^{\prime})
    \int dx \;
    \beta V(x)\rhoeqshort(x).
    \label{eq:deltaQ1-intermediate}
\end{align}
The second term in Eq.\,\eqref{eq:deltaQ1-intermediate} is zero due to 
Eq.\,\eqref{eq:normalization-of-rhoeq}, which implies 
$\partial_t\int dx \, \rhoeqshort(x) = \partial_t 1 = 0$.
The first term can be written in terms of the difference 
in entropy, $\Delta S^{eq}$, between $\rhoeqshort_{\boldsymbol{\lambda}(0)}(x)$
and $\rhoeqshort_{\boldsymbol{\lambda}(\Omega)}(x)$.
We recall the definition of the entropy $S^{eq}$ of an equilibrium distribution:
\begin{equation}
    S^{eq}_{\boldsymbol{\lambda}(t)} 
    = -\int dx\; \rhoeq(x)\log\rhoeq(x),
\end{equation}
the time derivative of which is $\int dx\; \beta V(x)\partial_t\rhoeqshort(x)$.
Thus we have
\begin{equation}
    \left<\Delta Q\right>_1 
    =
    - \frac{1}{\beta}\int_0^{\Omega} dt \; 
    \partial_t S^{eq}_{\boldsymbol{\lambda}(t)}
    =-\frac{1}{\beta}\Delta S^{eq}.
    \label{eq:q1}
\end{equation}
If we truncate the approximation of $\rho(x;t)$ at $\delta\rho^{(1)}(x;t)$, 
we reproduce the quasistatic 
Clausius equality for diffusive systems~\cite{hatano-sasa-2001,maes2014,mandal2016},
\begin{equation}
    \beta \langle \Delta Q\rangle \left[\boldsymbol{\Lambda}\right] +\Delta S^{eq}=0.
\end{equation}

The third contribution to $\langle\Delta Q\rangle$ is
\begin{equation}
    \left<\Delta Q\right>_2
    =\frac{1}{\gamma\beta^2}
    \int_0^{\Omega} dt \int dx \; 
    \delta\rho^{(2)}(x;t)
    \; e^{\beta V(x)}\partial^2_{x}e^{-\beta V(x)}.
\end{equation}
Similar to the calculation of $\left<\Delta Q\right>_1$, we use
Eq.\,\eqref{eq:deltarho2} to replace $\delta\rho^{(2)}$, apply Eq.\,\eqref{eq:trick1}, 
and then Eq.\,\eqref{eq:action-of-L-on-G}. This gives
\begin{align}
    \left<\Delta Q\right>_2
    = &-\frac{1}{\beta}\int_0^{\Omega} dt \; \bigg(
    \int dx^{\prime\prime}\;
    \partial_t\delta\rho^{(1)}(x^{\prime\prime};t)
    \beta V(x^{\prime\prime})
    \nonumber\\
    &~~~~~
    -\int dx^{\prime\prime}\;
    \partial_t\delta\rho^{(1)}(x^{\prime\prime};t)
    \int dx\; \rhoeqshort(x)\beta V(x)
    \bigg).
    \label{eq:deltaQ2-intermediate}
\end{align}
The second term in Eq.\,\eqref{eq:deltaQ2-intermediate} is zero
due to Eq.\,\eqref{eq:deltarho-integrates-to-zero}.
Writing $-\beta V(x^{\prime\prime}) = \log \rhoeqshort(x^{\prime\prime}) + \log Z$,
the first term can be rewritten as
\begin{align}
    \left<\Delta Q\right>_2
    &=
    -\frac{1}{\beta}\int_0^{\Omega} dt
    \int dx^{\prime\prime}\;
    \delta\rho^{(1)}(x^{\prime\prime};t)\;
    \partial_t\log\rhoeqshort(x^{\prime\prime})
    \nonumber\\
    &~~~
    -\frac{1}{\beta}\int_0^{\Omega} dt\;
    \left(\partial_t\log Z\right)
    \int dx^{\prime\prime}\;
    \delta\rho^{(1)}(x^{\prime\prime};t)
    \nonumber\\
    &~~~
    -
    \int_0^{\Omega} dt
    \int dx^{\prime\prime}\;
    \partial_t\left(
    \delta\rho^{(1)}(x^{\prime\prime};t)
    V(x^{\prime\prime})\right).
    \label{eq:deltaQ2-second-intermediate}
\end{align}
We evaluate the three terms in Eq.\,\eqref{eq:deltaQ2-second-intermediate} in reverse order.

The third term is the integral of a total time derivative and
depends only on the initial and final values of $\boldsymbol{\lambda}$ 
and $\dot{\boldsymbol{\lambda}}$.
It can be written as
\begin{equation}
    A(\boldsymbol{\lambda}(\Omega),\dot{\boldsymbol{\lambda}}(\Omega))
    -A(\boldsymbol{\lambda}(0),\dot{\boldsymbol{\lambda}}(0))
    \equiv \Delta A,
    \label{eq:q2-correction-term}
\end{equation}
where the function $A$ is given by
\begin{equation}
    A 
    =-\sum_{i}\dot{\lambda}_i\int\int dx \, dx^{\prime} \,
    \pot(x)\Green(x;x^{\prime})
    \frac{\partial\rhoeq}{\partial \lambda_i}(x^{\prime}).
    \label{eq:defn-of-A}
\end{equation}

The second term in Eq.\,\eqref{eq:deltaQ2-second-intermediate} evaluates to 
zero due to Eq.\,\eqref{eq:deltarho-integrates-to-zero}.

Lastly, the integral with respect to $x^{\prime\prime}$ in the first term in Eq.\,\eqref{eq:deltaQ2-second-intermediate}
can be rewritten as a quadratic form,
\begin{equation}
    -\int dx^{\prime\prime}\;
    \delta\rho^{(1)}(x^{\prime\prime};t)\;
    \partial_t\log\rhoeq(x^{\prime\prime})
    =
    \dot{\boldsymbol{\lambda}}^{\top}\boldsymbol{\zeta}\dot{\boldsymbol{\lambda}},
    \label{eq:q2}
\end{equation}
where the elements of the matrix $\metric$ are given by the formula
% \begin{widetext}
\begin{align}
    \zeta_{ij}=
    -\int\int &dx^{\prime} dx^{\prime\prime} \; 
    \bigg[\rhoeq(x^{\prime\prime}) 
    \left(\frac{\partial}{\partial \lambda_i} 
    \log \rhoeq(x^{\prime\prime})
    \right)\nonumber\\
    &\Green(x^{\prime}; x^{\prime\prime})
    \left(\frac{\partial}{\partial \lambda_j} 
    \log \rhoeq(x^{\prime})
    \right)\bigg].
    \label{eq:thermodynamic_metric}
\end{align}
% \end{widetext}

$\metric$ is clearly symmetric. We now prove that it is also positive definite.
In terms of $\psi_0$ and $\SchrGreen$, Eq.\,\eqref{eq:thermodynamic_metric}
takes the following simple form:
\begin{equation}
    \zeta_{ij}=2\int\int dx^{\prime} dx^{\prime\prime} \;
    \frac{\partial \psi_0(x^{\prime\prime})}{\partial \lambda_i}\,
    \SchrGreen(x^{\prime};x^{\prime\prime})\,
    \frac{\partial \psi_0(x^{\prime})}{\partial \lambda_j}.
    \label{eq:metric-Schr}
\end{equation}
Consider the quadratic form 
$\dot{\boldsymbol{\lambda}}^{\top}\boldsymbol{\zeta}\dot{\boldsymbol{\lambda}}$.
Using Eqs.\,\eqref{eq:greens-fn-of-schr} and \eqref{eq:alpha-and-E} in
Eq.\,\eqref{eq:metric-Schr}, we have
\begin{align}
    \dot{\boldsymbol{\lambda}}^{\top}
    \boldsymbol{\zeta}\dot{\boldsymbol{\lambda}}
    = 
    -\sum_{n\neq 0}\frac{1}{\alpha_n}
    \left(2\sum_{i=1}^{k}\int dx \;
    \dot{\lambda}_i
    \psi_n(x)
    \frac{\partial \psi_0}{\partial\lambda_i}(x)
    \right)^2
    > 0.
    \label{eq:proof-of-psd-metric}
\end{align}
The last inequality is due to the fact that $-\alpha_{n\neq 0}>0$.
Thus, the eigenvalues of $\metric$ are positive.
$\metric$ therefore induces a Riemannian metric on the space $\mathcal{C}$,
and can be identified as the thermodynamic metric \cite{sivak2012} for driven Brownian systems 
described by Eq.\,\eqref{eq:Fokker-Planck} with confining Schr\"{o}dinger potentials.
We note that Eq.\,\eqref{eq:thermodynamic_metric} contains all the timescales in the problem since $\Green$ contains a sum 
over all the eigenvalues of $\FPop$.

It becomes necessary now to distinguish between covariant and 
contravariant quantities; therefore, from this point onward in the discussion,
we will write control variables with raised indices, as
$\lambda^{i}$.

We can explicitly check that $\metric$ transforms correctly under a change of coordinates.
Using the representation given by Eq.\,\eqref{eq:metric-Schr}, it is simple to see that
under a continuous, invertible transformation (diffeomorphism)
$\boldsymbol{\lambda}\rightarrow\boldsymbol{\phi}(\boldsymbol{\lambda})$,
the elements of the new metric $\Tilde{\boldsymbol{\zeta}}(\boldsymbol{\phi})$ 
in $\boldsymbol{\phi}$ space are given by
\begin{equation}
    \Tilde{\zeta}_{kl}=\sum_{i,j}\zeta_{ij}\frac{\partial\lambda^i}{\partial \phi^k}\frac{\partial\lambda^j}{\partial \phi^l}.
\end{equation}
This transformation law for the metric holds due to the two partial derivatives
with respect to $\lambda^{i}$ and $\lambda^{j}$ in Eq.\,\eqref{eq:metric-Schr},
which in turn derive from the two partial derivatives with respect to time
in $\delta\rho^{(2)}(x;t)$. Therefore, even though  Eq.\,\eqref{eq:heat-production} 
has no geometric structure in general that 
we can discover, the specific form of $\delta\rho^{(2)}(x;t)$ introduces
geometric structure in the average dissipation.
We will see shortly that this emergent structure persists in Eq.\,\eqref{eq:heat-production} 
to all orders in $\nu$.

We emphasize
that Eq.\,\eqref{eq:thermodynamic_metric} is distinct from the formula for a thermodynamic
metric given in Eq.\,12 in \cite{sivak2012}, which was the first work to 
derive a thermodynamic metric for mesoscopic systems with time-varying relaxation times.
% The latter formula was the first generalization
% of thermodynamic length formalism to microscopic systems with time-varying relaxation times,
As mentioned previously, this formula was derived in the linear response
regime with a slow driving assumption. Evaluating it involves computing an integral with respect
to time over the linear response function, which is the average two-point time correlation
function of deviations of the conjugate forces from their equilibrium values.

Gathering the contributions from Eqs.\,\eqref{eq:q0}, \eqref{eq:q1},
\eqref{eq:q2-correction-term} and \eqref{eq:q2}, 
we have the following formula for the average heat up to terms of the order of
$\nu^2$ in Eq.\,\eqref{eq:solution-to-FP-equation}:
\begin{align}
    \beta\langle \Delta Q\rangle \left[\boldsymbol{\Lambda}\right]
    &= 
    0
    -\Delta S^{eq}
    +\int_0^{\Omega} dt \; \dot{\boldsymbol{\lambda}}(t)
        \metric
        \dot{\boldsymbol{\lambda}}(t)^{\top}
    +\beta \Delta A.
    \label{eq:derivationofmetric}
\end{align}

To minimize Eq.\,\eqref{eq:derivationofmetric} over protocols, we can define the action
\begin{equation}
    S[\boldsymbol{\lambda}(t)]
    =\beta\Delta A
    +
    2\int_0^{\Omega} dt \; \frac{1}{2}\dot{\boldsymbol{\lambda}}(t)
    \metric
    \dot{\boldsymbol{\lambda}}(t)^{\top}. 
    \label{eq:action}
\end{equation}
The equations of motion follow by setting the variation $\frac{\delta S}{\delta\lambda^i}$
of $S$ with respect to $\lambda^i$
to zero, subject to the constraints $\delta\lambda^i(0)=\delta\lambda^i(\Omega)=0$ $\forall i$.
These constraints imply $\delta A(0)=\delta A(\Omega)=0$, 
and therefore only the second 
term in Eq.\,\eqref{eq:action} contributes to the equations of motion.
These are the Euler-Lagrange equations of the Lagrangian $L=\frac{1}{2}\dot{\boldsymbol{\lambda}}^{\top}\boldsymbol{\zeta}\dot{\boldsymbol{\lambda}}$:
\begin{equation}
    \frac{d}{dt}\left(2\sum_{j}\zeta_{pj}\dot{\lambda}^j\right)
    =\sum_{i,j}\dot{\lambda}^i\frac{\partial \zeta_{ij}}{\partial \lambda^p}\dot{\lambda}^j,
    \,\, p\in(1,\dots, k).
    \label{eq:general-Euler-Lagrange}
\end{equation}
Opening out the time derivative on the left-hand side of 
Eq.\,\eqref{eq:general-Euler-Lagrange}, a straightforward calculation
shows that it is equivalent to
\begin{equation}
    \ddot{\lambda}^p+\sum_{i,j}\Gamma^{p}_{ij}\dot{\lambda}^i\dot{\lambda}^j=0,
    \,\, p\in(1,\dots, k),
    \label{eq:general-Euler-Lagrange-covariant-form}
\end{equation}
where $\Gamma^{p}_{ij}$ is the Christoffel symbol of the second kind,
\begin{equation}
    \Gamma^{p}_{ij}=\frac{1}{2}\sum_{m}\zeta^{pm}\left(
    \frac{\partial \zeta_{mi}}{\partial x^j}
    + \frac{\partial \zeta_{mj}}{\partial x^i}
    - \frac{\partial \zeta_{ij}}{\partial x^m}
    \right).
    \label{eq:Christoffel}
\end{equation}
Equations\,\eqref{eq:general-Euler-Lagrange-covariant-form} are also
the equations of motion of the Lagrangian 
$\Tilde{L}=\sqrt{\dot{\boldsymbol{\lambda}}\boldsymbol{\zeta}\dot{\boldsymbol{\lambda}}^{\top}}$
in the arc-length parametrization \cite{poisson_2004}.
In other words, these are geodesic equations of the control parameter space
$\mathcal{C}$.

Due to the spectral properties of $\Schr$,
Eq.\,\eqref{eq:proof-of-psd-metric} also indicates that the quadratic form
$\dot{\boldsymbol{\lambda}}^{\top}\boldsymbol{\zeta}\dot{\boldsymbol{\lambda}}$
is always finite.
Therefore, if $\pot$ is such that $\Schrpot$ is confining, 
and the perturbative expansion given by Eq.\,\eqref{eq:solution-to-FP-equation} holds
over the time period $\Omega$,
we are guaranteed that $\metric$ exists and is well defined over the course 
of driving. Then, up to terms of the order of $\nu^2$ in Eq.\,\eqref{eq:solution-to-FP-equation},
optimal protocols $\boldsymbol{\Lambda}^{\text{opt}}$
are geodesics
in $\mathcal{C}$ with respect to the 
length measure defined 
by $\metric$.

We note that in a specific optimal problem,
the invariance of the geodesic equations to reparametrizations
of $\mathcal{C}$ is broken by the boundary conditions, 
in which the \emph{identities} of the control parameters, 
along with their initial and final values, are specified.
For example, in the next section, we consider the harmonic potential 
$\pot(x)=\kappa x^2/2 + Ex$ with time-dependent electric field $E$ 
and spring constant $\kappa$. 
The choice of these two control parameters
breaks the diffeomorphism invariance
of Eq.\,\eqref{eq:general-Euler-Lagrange-covariant-form}
for this problem instance.

The diffeomorphism invariance of the geodesic equations
suggests that it is appropriate to write $\pot$ in such a way that all components
of $\boldsymbol{\lambda}$ have matching units. One way to do this is to 
introduce a fixed length scale $\ell$ and rescale $x$ as $x\rightarrow x/\ell$.
For example, in the harmonic potential defined previously, 
the control parameters $\kappa$ and $E$ have different units.
Rescaling $x$ by $\ell$, we can instead write 
$\pot(x/\ell)=(\ell^2\kappa) (x/\ell)^2/2 + (\ell E)x/\ell$. 
The new control vector is $\boldsymbol{\lambda}=(\ell^2\kappa, \ell E)$,
both components of which have 
%the unit 
units
of energy. Applying diffeomorphisms
that may scramble the two control parameters now makes sense.
We can choose $\ell$ to be such that $\beta \ell E=1$ or, equivalently,
such that $\beta \ell^2\kappa=1$.

We end this section with a note on higher-order terms in the average heat production.
By calculations 
analogous to those for
$\langle \Delta Q\rangle_2$, it is 
straightforward
to
establish that for any $w\geq 2$, the contribution to Eq.\,\ref{eq:heat-production}
from $\delta\rho^{(w)}(x;t)$ takes the form
\begin{equation}
    \beta\langle \Delta Q\rangle_w = 
    \beta\Delta A_{w} 
    + \int_0^{\Omega} dt \; \sum_{i_1,\dots,i_w} \dot{\lambda}^{i_1}\dots\dot{\lambda}^{i_w}\Xi^{(w)}_{i_1\dots i_w},
    \label{eq:higher-order-terms-in-Q}
\end{equation}
where $A_w$ is a term that depends only on the values of $\boldsymbol{\lambda}$ and
$\dot{\boldsymbol{\lambda}}$ at times $0$ and $\Omega$, and $\boldsymbol{\Xi}^{(w)}$ is
an object with $w$ indices.
[In the notation of Eq.\,\eqref{eq:higher-order-terms-in-Q},
the quantity $A$ defined in Eq.\,\eqref{eq:defn-of-A} is $A_2$, and
the thermodynamic metric $\boldsymbol{\zeta}$ is $\boldsymbol{\Xi}^{(2)}$.]
Due to the fact that $\delta\rho^{(w)}(x;t)$ contains exactly $w$
derivatives with respect to time, under a reparametrization 
$\boldsymbol{\lambda}\rightarrow\boldsymbol{\phi}(\boldsymbol{\lambda})$,
$\boldsymbol{\Xi}^{(w)}$ obeys the
transformation law $\Tilde{\Xi}^{(w)}_{j_1\dots j_w}=\sum_{i_1,\dots, i_w}\Xi^{(w)}_{i_1\dots i_w}\partial_{\phi^{j_1}}\lambda^{i_1}\dots \partial_{\phi^{j_w}}\lambda^{i_w}$, 
and is therefore a rank-$w$ tensor.
Thus, if the conditions for the existence of Eq.\,\eqref{eq:solution-to-FP-equation} 
are met, geometric structure is emergent in Eq.\,\eqref{eq:heat-production}
at all orders in $\nu$.

Up to terms of the order of $\nu^{k}$ in $\rho(x;t)$,
the Lagrangian of the optimal control problem is given by $L^{(w)}=\sum_{w=2}^{k}\sum_{i_1,\dots,i_w} \dot{\lambda}^{i_1}\dots\dot{\lambda}^{i_w}\Xi^{(w)}_{i_1\dots i_w}$;
like Eq.\,\eqref{eq:defn-of-A}, the 
$\Delta A_w$ for $w\geq 3$ do not participate in the Euler-Lagrange equations
for $\boldsymbol{\Lambda}^{\text{opt}}$.
Predictions of optimal protocols can be refined beyond the solutions of Eq.\,\eqref{eq:general-Euler-Lagrange-covariant-form} by including 
terms of the order of $w=3$ and higher in $L^{(w)}$.
The $\boldsymbol{\Xi}^{(w)}$---and therefore $L^{(w)}$---can easily be expressed in terms of $\rho_{\boldsymbol{\lambda}(t)}^{eq}$ and $\Green$.
For example, the elements of $\boldsymbol{\Xi}^{(3)}$ are given by
\begin{align}
    \Xi^{(3)}_{ijk} &= -\int\int dx dx^{\prime} \; 
    \frac{\partial\log\rhoeq(x)}{\partial \lambda^i}
    \Green(x;x^{\prime})\nonumber\\
    &\frac{\partial}{\partial\lambda^j}\bigg(
    \int dx^{\prime\prime} \;
    \Green(x^{\prime};x^{\prime\prime})\rhoeq(x^{\prime\prime})
    \frac{\partial\log\rhoeq(x^{\prime\prime})}{\partial\lambda^k}\bigg).
\end{align}
We leave the study of possible interpretations of $\boldsymbol{\Xi}^{(w)}$ for 
$w\geq 3$ and the development of solutions of the Euler-Lagrange equations of 
$L^{(w)}$ for $w\geq 3$ to future work.

\subsection{Relationship of $\zeta$ to previously proposed formula for a thermodynamic metric}

In \cite{zulkowski2015-optcontrol-overdamped}, the authors propose an 
approximate formula for a thermodynamic metric involving only $\rhoeq$.
Call this metric $\boldsymbol{\chi}$. 
Using the notation $\pieq$ to refer to the cumulative distribution function
\begin{equation}
    \pieq(x) = \int_{-\infty}^{x} dx^{\prime}\rhoeq(x^{\prime}),
\end{equation}
the elements of $\boldsymbol{\chi}$ are given by
\begin{equation}
    \chi_{ij} = \int dx \;
    \frac{\gamma\beta}{\rhoeq(x)}
    \left(\frac{\partial}{\partial\lambda_i}\pieq(x)\right)\left(\frac{\partial}{\partial\lambda_j}\pieq(x)\right).
    \label{eq:p-metric}
\end{equation}
The advantage of this formula is that it is entirely local in $x$, depending only on $\rhoeq$
and not on $\Green$, which is nonlocal in $x$ and contains all the natural timescales of the system.
In the case of a harmonic potential, it can be checked by explicit calculation that
$\boldsymbol{\zeta}$ and $\boldsymbol{\chi}$ are identical. For more general potentials, we
show that in a certain limit, Eq.\,\eqref{eq:thermodynamic_metric}
can be written as Eq.\,\eqref{eq:p-metric} plus correction terms.

For this part of the discussion only, we restrict ourselves to potentials of the form
\begin{equation}
    \pot(x)=g(x)+\sum_{i=1}^{m}a_ix^i,
    \label{eq:polynomial-potential}
\end{equation}
where $m\geq 4$ is even, and $a_m>0$. The $a_i$ are functions of $\boldsymbol{\lambda}(t)$.
$g(x)$ is any function of $x$ and $\boldsymbol{\lambda}$ that is finite in the limit $|x|\rightarrow\infty$.
At large $x$, this potential is dominated by the $x^m$ term.
In fact, it contains a natural length scale 
$x_0$ defined as the value of $x$ at which the ratio $\pot(x_0)/a_mx_0^m$
is of the order of $1$.
For such a potential, it is the case that 
\begin{equation}
    \lim_{|x|\rightarrow\infty}\,e^{\beta \pot(x)/2}
    \frac{\partial}{\partial \lambda_i}\pieq(x) = 0,
\end{equation}
and integrals over $x$ of the quantity in the limit converge.
This can be established using the asymptotic expansion of $1-\pieq(x_0)\sim\int_{x_0}^{\infty} dy \, e^{-\beta a_my^m}$:
\begin{equation}
    \int_{x_0}^{\infty} dy \, e^{-\beta a_my^m}
    \approx\frac{e^{-\beta a_mx_0^m}}{x_0^{m-1}}\left(1+\mathcal{O}\left(\frac{1}{x_0}\right)\right).
    \label{eq:asymptotic-expansion}
\end{equation}
The first term in the expansion can be verified by differentiating both sides of
Eq.\eqref{eq:asymptotic-expansion} with respect to $x_0$.

In the following, we drop the subscript $\boldsymbol{\lambda}(t)$ for brevity.
We use the notation $\zeta_{ij}^{x_0}$ and $\chi_{ij}^{x_0}$ to denote
Eqs.\,\eqref{eq:thermodynamic_metric} and \eqref{eq:p-metric} with all integrals
evaluated between $-x_0$ and $x_0$.

Using 
$\partial_x\Pi^{eq}(x)=\rhoeqshort(x)$,
Eq.\,\eqref{eq:thermodynamic_metric} can be rewritten as
\begin{align}
    \zeta^{x_0}_{ij}=
    -\int_{-x_0}^{x_0} dx^{\prime} dx^{\prime\prime} \; 
    &\frac{\partial^2 \Pi^{eq}(x^{\prime\prime})}{\partial\lambda_i \partial x^{\prime\prime}}\;
    \frac{G(x^{\prime}; x^{\prime\prime})}{\rhoeq(x^{\prime})}\;
    \frac{\partial^2 \Pi^{eq}(x^{\prime})}{\partial\lambda_j \partial x^{\prime}}.
\end{align}
Integrating by parts twice, this is
\begin{align}
    \zeta^{x_0}_{ij}=
    -\int_{-x_0}^{x_0} dx^{\prime} dx^{\prime\prime} \; 
    \frac{\partial \Pi^{eq}(x^{\prime\prime})}{\partial\lambda_i} \;
    \Theta(x^{\prime},x^{\prime\prime}) \;
    \frac{\partial \Pi^{eq}(x^{\prime})}{\partial\lambda_j},
    \label{eq:metric-comparison-intermediate}
\end{align}
where
\begin{equation}
    \Theta(x^{\prime},x^{\prime\prime})
    =\frac{\partial^2}{\partial x^{\prime}\partial x^{\prime\prime}}
    \frac{G(x^{\prime};x^{\prime\prime})}{\rhoeqshort(x^{\prime})}.
    \label{eq:Theta}
\end{equation}
For potentials of the form given by Eq.\,\eqref{eq:polynomial-potential}, 
the boundary terms in Eq.\,\eqref{eq:metric-comparison-intermediate} 
are exponentially suppressed in $x_0$, that is, they are of the order of $e^{-\beta a_mx_0^m}$.
Opening out the derivatives in $\Theta$, we find that it satisfies the differential
equation
\begin{equation}
    \frac{1}{\gamma\beta}\frac{\partial}{\partial x^{\prime}}\rhoeqshort(x^{\prime})\Theta(x^{\prime},x^{\prime\prime})
    =\FPopshort(x^{\prime})\frac{\partial G(x^{\prime}; x^{\prime\prime})}{\partial x^{\prime\prime}}.
\end{equation}
Applying Eq.\,\eqref{eq:action-of-L-on-G}, this is
\begin{equation}
    \frac{\partial}{\partial x^{\prime}}
    \left(\rhoeqshort(x^{\prime})\Theta(x^{\prime},x^{\prime\prime})
    +\gamma\beta\delta\left(x^{\prime}-x^{\prime\prime}\right)
    \right)
    =0.
\end{equation}
The solution to this differential equation 
is a family of functions $h_{x^{\prime}}(x^{\prime\prime})$
parameterized by $x^{\prime}$. 
We choose to work with $h$ evaluated at 
$x^{\prime}=x_0$, henceforth notated simply as $h(x^{\prime\prime})$:
\begin{equation}
    h(x^{\prime\prime})=\rhoeqshort(x_0)\Theta(x_0,x^{\prime\prime})+\gamma\beta\;
    \delta(x_0-x^{\prime\prime}).
\end{equation}
In terms of $h$, Eq.\,\eqref{eq:Theta} can be written as
\begin{equation}
    \Theta(x^{\prime},x^{\prime\prime}) = 
    \frac{1}{\rhoeqshort(x^{\prime})}
    \left(
    -\beta\gamma \;\delta(x^{\prime}-x^{\prime\prime}) +h(x^{\prime\prime})\right).
\end{equation}
Substituting this in Eq.\,\eqref{eq:metric-comparison-intermediate}, we find
\begin{equation}
    \zeta^{x_0}_{ij}=\chi^{x_0}_{ij} +\Delta^{x_0}_{ij},
    \label{eq:metric-comparison}
\end{equation}
where 
\begin{equation}
    \Delta^{x_0}_{ij}=- \int_{-x_0}^{x_0} dx^{\prime} dx^{\prime\prime} \; \frac{\beta\gamma}{\rhoeqshort(x^{\prime})}
    \frac{\partial \Pi^{eq}(x^{\prime})}{\partial \lambda_j}
    h(x^{\prime\prime})
    \frac{\partial \Pi^{eq}(x^{\prime\prime})}{\partial \lambda_i}.
\end{equation}
Once again using the asymptotic expansion given by Eq.\,\eqref{eq:asymptotic-expansion},
it can be shown that $\Delta^{x_0}_{ij}$ is of the order of $e^{-\beta a_mx_0^m}$. 
We note that it is necessary to evaluate the function $h_{x^{\prime}}$ at $x^{\prime}\geq x_0$ 
to arrive at this conclusion; otherwise it is not clear how to 
estimate the size of $\Delta^{x_0}_{ij}$.
Therefore, we finally arrive at
\begin{equation}
    \zeta^{x_0}_{ij}=\chi^{x_0}_{ij} +\mathcal{O}(e^{-\beta a_mx_0^m}).
    \label{eq:metric-comparison-final}
\end{equation}

From Eq.\,\eqref{eq:metric-comparison-final}, we see that in the limit $|x_0|\rightarrow\infty$, 
all correction terms go to zero, and we have $\zeta_{ij}-\chi_{ij}\rightarrow 0$. 
However, this limit is not physically valid---it is simple to check that
as $x_0\rightarrow\infty$, Eq.\eqref{eq:Fokker-Planck} is trivialized to $0=0$.
Thus, for general potentials, we cannot expect the two formulas $\boldsymbol{\zeta}$ 
and $\boldsymbol{\chi}$ to be equivalent.
As previously mentioned, the quadratic potential is an interesting exception %in 
for which 
it can be explicitly checked that both $\boldsymbol{\zeta}$ and $\boldsymbol{\chi}$ 
evaluate to the same quantity.

The calculation leading to Eq.\,\eqref{eq:metric-comparison} is a proof of the 
formula given by Eq.\,\eqref{eq:p-metric} for polynomial potentials. In 
\cite{zulkowski2015-optcontrol-overdamped}, the class of potentials
for which Eq.\,\eqref{eq:p-metric} converges was not established.
We further note that we expect a relation similar to Eq.\,\eqref{eq:metric-comparison-final}
to hold for potentials that grow faster than Eq.\,\eqref{eq:polynomial-potential};
%, 
for example, 
%for 
$V(x)=e^{b|x|}$ with $b>0$. The specifics of the asymptotic analysis proving this point will differ
from what is presented here.

\section{The harmonic oscillator in an electric field}

We calculate $\boldsymbol{\zeta}$ for a one-dimensional system 
of charge $q$
in a harmonic potential with time-dependent spring constant 
$\kappa(t)$ and subject to an external electric field $E(t)$.
The control vector is $\boldsymbol{\lambda}(t) = (\kappa(t), E(t))$,
where $\kappa>0$ and $E\in\mathbb{R}$.
The potential is
\begin{equation}
    \pot(x) 
    = \frac{1}{2}\kappa x^2-qEx
    = \frac{1}{2}\kappa\left(x-\theta\right)^2 - \frac{\kappa}{2}\theta^2.
    \label{eq:harmonic-oscillator-potential}
\end{equation}
In the second equality we have defined the new variable $\theta = E/\kappa$. 
The electric field can be interpreted as an offset in the center of the harmonic trap.

The Fokker-Planck operator 
for this system is
\begin{equation}
    \FPop(x) 
    = \frac{1}{\gamma}\left[\kappa(t) 
    + \kappa(t)\left(x-\theta(t)\right)\frac{\partial}{\partial x} 
    + \frac{1}{\beta}\frac{\partial^2}{\partial x^2}\right].
    \label{eq:FP_operator_harmonic}
\end{equation}
The eigenfunctions $\psi_n$ of the corresponding Schr\"{o}dinger operator are
given by the Hermite functions ~\cite{risken1984}. 
Using $H_n$ to denote the $n^{th}$ Hermite polynomial, 
the right and left eigenfunctions are
\begin{subequations}
\begin{align}
    \rho_{r,n}(x) 
    &= \frac{1}{\sqrt{2^n n!}}
    \sqrt{\frac{\beta\kappa}{2\pi}}
    e^{-\frac{1}{2} \beta\kappa(x-\theta)^2}
    H_n\left(\sqrt{\frac{\beta\kappa}{2}}(x-\theta)\right),
    \\
    \rho_{l,n}(x) 
    &= \frac{1}{\sqrt{2^n n!}}
    H_n\left(\sqrt{\frac{\beta\kappa}{2}}(x-\theta)\right).
\end{align}
\end{subequations}
The corresponding eigenvalues are $-\kappa n/\gamma$.
The equilibrium distribution at any 
given time $t$ is a 
normalized Gaussian distribution 
with mean $\theta$ and variance $1/\beta\kappa$,
\begin{equation}
    \rhoeq(x) = \sqrt{\frac{\beta\kappa}{2\pi}}e^{-\frac{1}{2}\beta\kappa (x-\theta)^2}.
\end{equation}

We proceed to 
calculate the four elements, beginning with 
$\zeta_{11} = \zeta_{\kappa\kappa}$:
\begin{widetext}
\begin{equation}
\begin{aligned}
    \zeta_{\kappa\kappa} 
    = -&\int dx \int dy \,\sqrt{\frac{\beta\kappa}{2\pi}}
    e^{-\frac{1}{2}\beta\kappa (y-\theta)^2}
    \left(\frac{1}{2\kappa}-\frac{\beta(x-\theta)^2}{2}\right)
    \left(\frac{1}{2\kappa}-\frac{\beta(y-\theta)^2}{2}\right)
    \\
    & \sum_{n \neq 0}
    -\frac{\gamma}{\kappa n}\frac{1}{2^n n!}
    \sqrt{\frac{\beta\kappa}{2\pi}}
    e^{-\frac{1}{2}\beta\kappa (x-\theta)^2}
    H_n \left(\sqrt{\frac{\beta\kappa}{2}}(x-\theta)\right)
    H_n \left(\sqrt{\frac{\beta\kappa}{2}}(y-\theta)\right).
\end{aligned}
\end{equation}
\end{widetext}
Transforming to the variables 
$x^\prime = \sqrt{\beta\kappa/2}(x-\theta),y^\prime = \sqrt{\beta\kappa/2}(y-\theta)$,
and using $\frac{1}{2}-x^{\prime 2} = -\frac{1}{4}H_2(x^{\prime})$, this is
\begin{equation}
    \zeta_{\kappa\kappa} = \frac{1}{\pi}\frac{\gamma}{\kappa^3} \frac{1}{16}
    \sum_{n \neq 0}\frac{1}{n 2^n n!}
    \left(
    \int dx^\prime
    \, e^{-x^{\prime 2}}
    H_2(x^{\prime})
    H_n\left(x^{\prime}\right)\right)^2.
\end{equation}
Applying the orthogonality property
\begin{equation}
    \int dx^\prime
    \, e^{-x^{\prime 2}}
    H_m(x^{\prime})
    H_n\left(x^{\prime}\right)
    = \delta_{mn}2^{n}n!\sqrt{\pi},
\end{equation}
we have
\begin{equation}
    \zeta_{\kappa\kappa} = \frac{\gamma}{4\kappa^3}.
\end{equation}
Similarly, the elements $\zeta_{\theta\kappa}$ and $\zeta_{\kappa\theta}$ are proportional
to the product
\begin{equation}
    \int dx^\prime
    \, e^{-x^{\prime 2}}
    \frac{1}{4}H_2(x^{\prime})
    H_n(x^{\prime})
    \int dy^\prime
    \, e^{-y^{\prime 2}}
    \frac{1}{2}H_1(y^{\prime})
    H_n(y^{\prime}),
\end{equation}
which evaluates to zero for all $n$.
Finally,
\begin{equation}
    \zeta_{\theta\theta}
    = \frac{2\beta\gamma}{\pi}
    \sum_{n \neq 0}\frac{1}{n 2^n n!}
    \left(
    \int dx^{\prime} \,
    \frac{1}{2}H_1(x^{\prime})
    H_n(x^{\prime})
    \right)^2
    =\beta\gamma.
\end{equation}

Gathering elements, we have
\begin{equation}
    \zeta = \gamma
    \begin{bmatrix}
    \left(4\kappa^3\right)^{-1} & 0
    \vspace{0.05in}\\
    0 & \beta
    \end{bmatrix}.
    \label{eq:oscillatormetric}
\end{equation}
As mentioned in the previous section, the same result is obtained by evaluating
Eq.\,\eqref{eq:p-metric} for this system. 
Equation\,\eqref{eq:oscillatormetric} is also identical to the 
result obtained by evaluating the formula for a thermodynamic metric
given in \cite{sivak2012} for a harmonic potential with time-varying spring
constant and trap center.

We can now calculate optimal protocols for the harmonic oscillator.
For the metric given by Eq.\,\eqref{eq:oscillatormetric}, 
Eq.\,\eqref{eq:q2} takes the form
\begin{equation}
    \int_{0}^{\Omega} dt \; \gamma\left(\frac{\dot{\kappa}^2}{4\kappa^3}+\beta\dot{\theta}^2\right)
    =\int_{0}^{\Omega} dt \; \gamma\left(\dot{\mu}^2+\beta\dot{\theta}^2\right).
    \label{eq:harmonic-oscillator-action}
\end{equation}
In the second equality above we have made the change of variables $\mu=\sqrt{\kappa}$.
This is a diffeomorphism for $\kappa > 0$.
From Eq.\,\eqref{eq:harmonic-oscillator-action} it is clear that the potential given by Eq.\,\eqref{eq:harmonic-oscillator-potential}
gives rise to a flat geometry in $(\mu, \theta)$ space.
However, the protocols have a nontrivial structure in the physical control parameter space $(\kappa, \theta)$ due to the existence of the forbidden region $\kappa\leq 0$.
The Euler-Lagrange equations corresponding to Eq.\,\eqref{eq:harmonic-oscillator-action} 
are $\ddot{\mu}=\ddot{\theta}=0$.
The solutions are straight lines in the $(\mu,\theta)$ plane.
Given initial and final values of the physical parameters---$\kappa_{\Omega}$ and $\kappa_{0}$, and similarly
for $\theta$---the protocol that minimizes
Eq.\,\eqref{eq:q2} is
\begin{subequations}\label{eq:optimal-HO-protocol}
\begin{align}
    &\theta^{\text{opt}}(t) = \frac{\theta_{\Omega}-\theta_{0}}{\Omega}t+\theta_0\\
    &\kappa^{\text{opt}}(t) = \left(\frac{\sqrt{\kappa_{\Omega}}-\sqrt{\kappa_{0}}}{\Omega}t+\sqrt{\kappa_{0}}\right)^2.
\end{align}
\end{subequations}
The optimal protocol demands a constant rate of change for $\theta$ and $\sqrt{\kappa}$.

In this example, we can explicitly check the consistency conditions of
Sec.\,\ref{sec:expansion-parameter}. To do so, it is convenient to rescale
the optimal control problem so that all control parameters are dimensionless.
This is easily done by first rescaling $x\rightarrow x/\ell$ where the length
measure $\ell$ is defined by $\beta\ell^2\kappa=1=E\ell\beta$,
as discussed at the end of Sec.\,\ref{sec:derivation-of-metric},
and then multiplying the potential (Eq.\,\eqref{eq:harmonic-oscillator-potential})
by $\beta$.
These rescalings do not disturb the optimal control problem.
We have the following optimal protocols for the dimensionless
control parameters 
$\left(\Tilde{\mu},\Tilde{\theta}\right)=\left(\sqrt{\beta\kappa\ell^2}, E/\kappa\ell\right)$:
\begin{subequations}\label{eq:optimal-HO-protocol-dimless}
\begin{align}
    &\Tilde{\theta}^{\text{opt}}(t) = \frac{\Tilde{\theta}_{\Omega}-\Tilde{\theta}_{0}}{\Omega}t+\Tilde{\theta}_0\\
    &\Tilde{\mu}^{\text{opt}}(t) = \frac{\Tilde{\mu}_{\Omega}-\Tilde{\mu}_{0}}{\Omega}t+\Tilde{\mu}_{0}.
\end{align}
\end{subequations}
These are of precisely the same form as Eq.\,\eqref{eq:optimal-HO-protocol}.
In terms of the dimensionless control parameters, the eigenvalues of 
the Fokker-Planck operator for the harmonic oscillator are given by 
$-\Tilde{\mu}^2n/\beta\ell^2\gamma$.
Therefore, under the optimal protocol,
the relaxation time of the Brownian system is
$\tau_{\alpha_1}=\beta\ell^2\gamma/\left(\Tilde{\mu}^{\text{opt}}\right)^2$.

Without loss of generality, we can assume $\dot{\Tilde{\mu}}^{\text{opt}}\geq\dot{\Tilde{\theta}}^{\text{opt}}$.
For ease of notation in what follows, we write the difference 
$\Tilde{\mu}_{\Omega}-\Tilde{\mu}_{0}$ as $\Delta\Tilde{\mu}$.
The longest driving timescale set by the optimal protocol is then
given by $\tau_{\lambda}=1/\dot{\Tilde{\mu}}^{\text{opt}}=\Omega/\Delta\Tilde{\mu}$.

Therefore, we have
\begin{equation}\label{eq:SHO-nu}
    \nu = \frac{\tau_{\alpha_1}}{\tau_{\lambda}}
    =\frac{\beta\ell^2\gamma}{\left(\Tilde{\mu}^{\text{opt}}\right)^2}\frac{\Delta\Tilde{\mu}}{\Omega}=\mathcal{O}\left(\frac{1}{\Omega}\right).
\end{equation}
$\nu$ can be made small by choosing $\Omega$, the duration of the protocol,
to be sufficiently long.

From Eq.\,\eqref{eq:optimal-HO-protocol-dimless}, 
we see that $\dot{\Tilde{\mu}}^{\text{opt}}$ is of the order of $1/\Omega$.
The rate of change of the spectrum of the Fokker-Planck operator also 
goes as $1/\Omega$. To see this, note that $\left|\dot{\alpha}_1\right|
=1/\tau_{\alpha_1}$. Differentiating this with respect to time, we find
$\left|\dot{\alpha}_1\right|=2\dot{\Tilde{\mu}}^{\text{opt}}\Tilde{\mu}^{\text{opt}}/\beta\ell^2\gamma = \mathcal{O}(1/\Omega)$ since $\dot{\Tilde{\mu}}^{\text{opt}}$
is $\mathcal{O}(1/\Omega)$.
Thus, both the control parameters and the spectrum of the Fokker-Planck
operator vary appreciably only on the timescale of the control parameters, 
and are roughly constant on the timescale of the system if $\Omega$ is chosen 
to be large.

Lastly, differentiating Eq.\,\eqref{eq:SHO-nu} with respect to time, 
we find that $\dot{\nu}$ is of the order of $1/\Omega^2$, i.e., $\mathcal{O}(\nu^2)$,
and is therefore suppressed on the control timescale.

\section{Summary and Future work}

We have developed a precise perturbative solution to Eq.\,\eqref{eq:Fokker-Planck}
and used it to calculate the heat generated in the environment when the 
external parameters of a small stochastic system are varied in time.
In so doing, we derived a new formula for the thermodynamic metric and all
correction terms at the same order in the perturbation theory. 

Both \cite{sivak2012} and \cite{zulkowski2015-optcontrol-overdamped} propose formulas
for thermodynamic metrics but do not establish the class of potentials for which those
formulas are valid.
The formula we have derived, given by Eq.\,\eqref{eq:thermodynamic_metric},
holds for potentials $\pot$ such that both $\pot$ and the associated Schr\"{o}dinger
potential $\Schrpot$ are confining. 
We have shown that for a subset of such potentials,
namely, those in Eq.\,\eqref{eq:polynomial-potential}, the formula given by Eq.\eqref{eq:p-metric}
of \cite{zulkowski2015-optcontrol-overdamped} is approximately valid.

We found that the expansion in $\nu$ has an emergent local diffeomorphism 
symmetry not present in the original formula, given by Eq.\eqref{eq:heat-production}, for average heat production. 
Every term of this expansion transforms as a tensor of this diffeomorphism symmetry. 
Restricting to the symmetric 2-tensor (metric) in the expansion, we explicitly worked out the 
equations for an optimal protocol. 
These equations of motion describe geodesics in the space of control parameters.

In future work, it would be interesting to study the
physical interpretation of the tensors $\boldsymbol{\Xi}^{(w)}$ for $w\geq 3$,
and to develop methods of calculating $\boldsymbol{\Lambda}^{\text{opt}}$
when these tensors are retained in the Lagrangian.
Additional directions for future research include extending the perturbation 
theory to underdamped systems and to higher spatial dimensions. 
For the latter, much of the formalism developed here will be applicable, but
it will be necessary to study the spectral properties of the Schr\"{o}dinger operator in higher dimensions.

In this paper, we derived a formula for the thermodynamic metric corresponding 
to the confining potential $\Schrpot$. This invites the following question: given a metric, 
what is the class of potentials that give rise to it? This may be especially interesting and 
 tractable in the case of two-dimensional Riemannian geometries.

\section{Acknowledgements}

The authors thank Dibyendu Mandal for collaboration on an earlier
version of this paper,
and thank Michael Y-S. Fang, Jeffrey M. Epstein, 
and Satya Majumdar for feedback on the manuscript. 
N.S.W. was supported by a Google Ph.D. Fellowship. 
R.V.Z. was supported by a National Science Foundation Graduate Research Fellowship under Grant No. DGE 1752814. 
M.R.D was supported in part by the U. S. Army Research Laboratory
and the U. S. Army Research Office under Contract No.
W911NF-20-1-0151.

\bibliographystyle{apsrev4-1}
\bibliography{main}

\end{document}